\documentclass[iop]{emulateapj}

\usepackage{amsmath}
\usepackage{color}

\newcommand{\oversim}[2]{\protect{\mbox{\lower0.5ex\vbox{%
   \baselineskip=0pt\lineskip=0.2ex
   \ialign{$\mathsurround=0pt #1\hfil##\hfil$\crcr#2\crcr\sim\crcr}}}}} 
\newcommand{\simgreat}{\mbox{$\,\mathrel{\mathpalette\oversim>}\,$}} 

\slugcomment{}

\newcommand{\be}{\begin{equation}}
\newcommand{\ee}{\end{equation}}

\shorttitle{Bulge--TDG correlation, {\it ApJ, in press}}
\shortauthors{Mart\'\i n L\'opez-Corredoira  \& Pavel Kroupa}

\begin{document}

\title{The number of tidal dwarf satellite galaxies in dependence of bulge index}
\author{Mart\'\i n L\'opez-Corredoira \altaffilmark{1,2} and Pavel Kroupa\altaffilmark{3}}
\altaffiltext{1}{Instituto de Astrof\'\i sica de Canarias, E-38205 La Laguna, Tenerife, Spain; martinlc@iac.es}
\altaffiltext{2}{Departamento de Astrof\'\i sica, Universidad de La Laguna, E-38206 La Laguna, Tenerife, Spain}
\altaffiltext{3}{Helmholtz-Institut f\"ur Strahlen- und Kernphysik,
  Universit\"at Bonn,  Nussallee 14-16, 53115 Bonn, Germany; pavel@astro.uni-bonn.de}

\begin{abstract}

We show that a significant correlation (up to 5$\sigma$) 
emerges between the bulge index, defined to be larger for larger 
bulge/disk ratio, in spiral galaxies with similar luminosities in 
the Galaxy Zoo 2 of SDSS and the number of
tidal-dwarf galaxies in the catalogue by Kaviraj et al. (2012).

In the standard cold or warm dark-matter
cosmological models the number of satellite galaxies correlates with
the circular velocity of the dark matter host halo. In
generalized-gravity models without cold or warm dark matter such a
correlation does not exist, because host galaxies cannot capture
in-falling dwarf galaxies due to the absence of dark-matter-induced
dynamical friction. However, in such models a correlation is expected
to exist between the bulge mass and the number of satellite galaxies,
because bulges and tidal-dwarf satellite galaxies form in encounters
between host galaxies. This  is not predicted by
dark matter models in which bulge mass and the number
  of satellites are a priori uncorrelated because
higher bulge/disk ratios do not imply higher dark/luminous
ratios. Hence, our correlation reproduces the prediction of
  scenarios without dark matter, whereas an explanation is not found
  readily from the a priori predictions of the standard scenario with
  dark matter.  Further research is needed to explore whether some
  application of the standard theory may explain this correlation.

\end{abstract}

\keywords{galaxies: bulges; galaxies: dwarfs; galaxies: formation;
  galaxies: interactions; dark matter}

\section{Introduction}

According to the cold or warm dark matter based
  standard model of cosmology (SMoC) dark-matter halos merge to form
  more massive host halos. This occurs because initially hyperbolic
  relative encounters between halos, some carrying galaxies, are
  dissipative due to the dark matter halos becoming partially unbound
  and due to dynamical friction.  Each major galaxy (with stellar mass
  $M_* \simgreat 5\times 10^9\,M_\odot$) is therefore at the centre of
  a major dark matter host halo (dark matter mass within the virial
  radius $M_{\rm DM} \simgreat 10^{11}\,M_\odot$, either isolated or
  as part of a cluster of galaxies, see fig.~4 in Wu \& Kroupa 2015)
  and ought to have many hundreds of dark-matter sub-halo satellites
  of which a certain fraction contain dwarf galaxies (Moore et
  al. 1999; Klypin et al. 1999), at least in field galaxies and
  Local-Group equivalents. The fraction of satellite galaxies
depends on how baryonic processes interplay in early structure
formation (e.g. Cooper et al. 2010; Guo et al. 2015; Kroupa et
al. 2010 and references therein). In this model and at any time,
  a dark-matter host halo contains many sub-halos many of which are
  constantly decaying towards the centre through dynamical friction
  while new sub-halos enter, and this process slows down with
  cosmological time and is punctuated by a few major merger events.
The expected distribution of satellite galaxies is spheroidal and
approximately isotropic around the major galaxies, as
  detailed and mathematically thorough analysis of the distribution of
  sub-halos has been showing (Metz et al. 2007; Pawlowski et al. 2014). 
Particularly  important in this context is the demonstration by Pawlowski et
  al. (2015b) that when baryonic physics is taken into account then
  the spatial distribution of dark-matter-dominated satellite galaxies
  is not affected and remains indistinguishable from being isotropic.
The satellite galaxies have high dynamical mass-to-light ratios
because their potentials are dominated by dark matter in their inner
regions (e.g. Simon \& Geha 2007). The number of satellite galaxies is
predicted to increase monotonically with the mass,
i.e. the circular velocity, of the host dark matter halo (Moore et
al. 1999; Kroupa et al. 2010 and references therein; Klypin et al. 2011; 
Ishiyama et al. 2013).

A different model has been formulated according to which gravitation
is generalized such that it is scale-invariant in the regime when the
gravitational acceleration is smaller than Milgrom's constant
$a_0\approx 3.8\,$pc/Myr$^2$ (Milgrom 2009; Famaey \&
  McGaugh 2012; see also Kroupa 2015; for another approach to
generalized gravitation see Moffat 2006). This model is consistent
with the observation that most galaxies are simple
dynamical objects lacking the scars from major and minor mergers
(Disney et al. 2008). It implies that galaxies evolve largely in
isolation, whereas some experience interactions and rarely mergers
(Kroupa 2015). Dwarf galaxies which condensed from the expanding
Universe after the Big Bang grow through gas accretion to become the
present-day star-forming galaxies. This model is largely
unexplored\footnote{But this can now be alleviated since the
  publication of the publicly-available Phantom of Ramses (PoR) code
  for galaxy evolution and cosmological structure formation by
  L\"ughausen, Famaey \& Kroupa (2015).}, but even if structure
formation would proceed similarly to that of the dark-matter based
models, the incoming dwarf or major galaxy would move past the primary
galaxy on a hyperbolic orbit because dynamical friction on the
non-existing dark matter halos would not dissipate relative orbital
energy. Only in rare near-head-on collisions would galaxies be
captured (Toomre 1977). Some would interact strongly
and would merge after many orbits due to the lack of
dark-matter-particle-driven dynamical friction
(Combes \& Tiret 2010). This Milgromian model would therefore
predict few satellite galaxies, as there is no significant mechanism
to capture a dwarf galaxy within a few hundred kpc from a major galaxy
unless the dwarf is initially on a radial orbit in which case it is
likely to be destroyed due to the external field effect (Famaey \&
McGaugh 2012; Kroupa 2015; Wu \& Kroupa 2015). However, the encounters
between galaxies draw out long tidal arms which, when gas rich,
fragment forming populations of star clusters and dwarf galaxies which
are strongly correlated in phase-space (Tiret \& Combes 2008;
Pawlowski et al. 2011; Yang et al. 2014).  Although
such tidal dwarf galaxies (TDGs) are dark matter free (Bournaud 2010),
they feign dark matter domination if their internal motions are
interpreted for virialised systems with Newtonian dynamics (Kroupa
1997; McGaugh \& Milgrom 2013; Yang et al. 2014; Pawlowski, McGaugh \&
Jerjen 2015a).  The number of TDGs is expected to statistically scale
with a parameter which is a measure of the degree of encounters a
given host has experienced. Classical- and pseudo-bulges typically
form in a galaxy after it experiences a tidal perturbation. Therefore
a convenient parameter as a measure of how much the host galaxy was
perturbed in the past would be the relative bulge mass or bulge index
$B$ (see Section~\ref{sec:data} below).

Two implications thus arise from the above: 
\begin{description}
\vspace{-2mm}

\item 1. Dark matter models: 
\begin{description}
     \item 1.1~The satellite galaxies are spheroidally
  distributed around their hosts. 
     \item 1.2~The number of satellite galaxies
  scales with the rotation velocity of the host galaxy which is a
  measure for the mass of the dark matter halo. 
\end{description}

\item 2. Milgromian and generalized (non-dark-matter) gravitation
  models: 

\begin{description}
\item 2.1~The satellite galaxies stemming from one encounter are
  phase-space correlated TDGs. The existence of
    planes or disks of satellites are a necessary consequence of this model.
     \item 2.2~The number of TDGs
  scales with the bulge index.
\end{description}

\end{description}

According to point 1.2~the relevant measure of the number of satellite
galaxies, $N_{\rm S}$, is the rotational velocity $V_{\rm C}$ of the
host galaxy, while according to point~2.2 the bulge index, $B$, is the
relevant measure. Thus an observational assessment of which model is
relevant for describing reality is possible by investigating
correlations of $N_{\rm S}$ with $V_{\rm C}$ or with $B$. {\it By
  selecting galaxies with a comparable $V_{\rm C}$ and thus comparable
  baryonic mass because galaxies follow the baryonic Tully-Fischer
  relation (McGaugh \& Schombert 2015), $N_{\rm S}$ should correlate
  with $B$ in model~2 and not in model~1} (Kroupa 2015).

The observational data situation is such that major galaxies have few
satellites (the missing satellite problem of the SMoC, Klypin et
al. 1999; Moore et al. 1999), and for those for which sufficiently
good three-dimensional spatial data exist anisotropic,
phase-space-correlated satellite galaxy distributions are
ubiquitous\footnote{Citing from Chiboucas et al. (2013): {\it ``In
    review, in the few instances around nearby major galaxies where we
    have information, in every case there is evidence that gas poor
    companions lie in flattened distributions.''}}.  Indeed, the Milky
Way has an extremely pronounced, highly significant rotating
disk-of-satellite (DoS) or vast-polar structure (VPOS) which includes
all material (satellite galaxies, globular clusters and gas and
stellar streams) beyond about 20~kpc distance (Kroupa et al. 2005;
Metz et al. 2007, 2008, 2009; Pawlowski et al. 2012, 2014, 2015a).
Andromeda has a very thin great-plane of Andromeda (GPoA) made up of
half of all its satellite galaxies (Metz et al. 2007; Metz et
al. 2009; Ibata et al. 2013, 2014a).  Both the DoS/VPOS and the GPoA
are rotational structures. Furthermore (Pawlowski et al. 2013): (i)
Not only are the VPOS and the GPoA extremely pronounced and
incompatible with the dark-matter models, they are also mutually
correlated in that the GPoA points precisely at the Milky Way (MW) and
both the VPOS and the GPoA have spin vectors which point into a
similar direction. (ii) The entire Local Group has a highly
symmetrical structure around the Milky-Way--Andromeda axis which is,
by all counts of probability and by virtue of the thinness of the
non-satellite vast planar structures, impossible to obtain in
structure formation within the SMoC, in which the dwarf galaxies in
the Local Group have largely independent formation and infall
histories. The nearest other galaxy group (M81) also has a
significantly anisotropic satellite galaxy distribution (Chiboucas et
al. 2013), and a significant number of bright host galaxies have
rotational satellite populations (Ibata et al. 2014b, 2015).  The
disagreement of these anisotropic satellite distributions with the
dark-matter-based models has been, for the first time, pointed out by
Kroupa et al. (2005), a problem or failure of the dark matter models
which has to-date not been solved within the dark-matter models.

The interested reader is referred to several rebuttals (Metz et
al. 2007, 2009; Pawlowski et al 2014; Ibata et al. 2014a; Ibata et
al. 2015; Pawlowski et al. 2015b) to obtain more details on the
numerous but incorrect claims by the standard-dark-matter teams that
the observed satellite galaxy anisotropies are consistent with the
satellite distributions with the SMoC. A unifying problem with these
is that simplified and often inadequate plane-finding and
characterization algorithms are used thereby ignoring the
sophisticated standardized mathematical tools developed for rigorous
statistical assessment of phase-space correlations of data by the work
of Metz et al. and Pawlowski et al. (see sec.~5.1.3 in Kroupa 2015 for
elaboration on this point).  This discussion need not be repeated
here, but along similar lines recently Cautun et al. (2015: ``Planes
of satellite galaxies: when exceptions are the rule``) appear to
suggest that the VPOS and GPoA are as unlikely in the models and
therefore they are normally unlikely and thus not a problem for the
SMoC. We can point out some of the shortcomings already: Cautun et
al. merely seek 11~{\it or fewer} satellites in their models which are
as correlated as the observed satellites, while the VPOS contains more
than~20, and Cautun et al. add the Sculptor satellite without taking
into account that it is on a couter-rotating orbit {\it within} the
VPOS. The Cautun et al. analysis does not specify the PAndAS footprint
they applied, which may introduce biases, and they compare model GPoA
3D velocities with the observed (i.e. projected) 2D GPoA structure
thereby introducing unaccounted-for biases. Cautun et al. neither take
into account that the VPOS contains much more than only~11 or fewer
satellites, nor that the GPoA is heavily lop-sided towards the MW, and
their choice of the MW obscuring region ($\pm19.5\,$degrees,
i.e. 33~\% of the sky) is too restrictive not being backed up by
data. Finally, even if their analysis were to be assumed to be
correct, it implies that the MW system would in the 5~\% and the
Andromeda system would be in the 9~\% tails of their distributions,
such that the Local Group would constitute a 0.45~\% outlier. Thus,
without even taking into account the mutual correlation of the GPoA
and of the VPOS, their result leads to very significant disagreement
with the SMoC, which they essentially appear to interpret to mean that
the Local Group properties are consistent with the SMoC. Another claim
based on simplified plane-searching strategies that the planar
satellite distributions are readily compatible with the dark-matter
model, has been made by Buck, Dutton \& Macci{\`o} (2015: ``The Plane
Truth: Andromeda analog thin Planes of Satellites are not kinematical
coherent structures''): They use newly discovered ``satellites'' of
Andromeda which are, however, beyond its virial radius. Buck et
al. misrepresent the literature (Lynden-Bell 1976 only knew of
6~satellite galaxies rather than the claimed~11; they ignore that
Kroupa et al. 2005 were the first to demonstrate that the distribution
of the 11~classical satellites are inconsistent with the SMoC), and it
is unclear what ``perfectly describes the collapse of dark matter
halos'' (their introduction) means. The analysis of the orbital pole
directions is flawed by them being mirrored onto one hemisphere, and
by them preselecting those satellites from 30 models with
Andromeda-like orbital poles to then argue that this selected ensemble
is similar to the Milky Way. Buck et al. argue that the planar
satellite arrangements of Andromeda and of the MW are both chance
occurrences which exist only for a few hundred million years, but they
neither elaborate why we would be observing them at this special epoch
simultaneously and mutually correlated, nor how their ``solution''
compares with the other solutions proposed by the dark-matter
community. There can be only one physical solution to the real Local
Group rather than many partially mutually excluding ones.

With this contribution we address the other-than-anisotropy
avenue of testing the fundamental models of physics concerning
cosmology, namely the number-of-satellites ($N_{\rm S}$) vs
host-galaxy property.  Kroupa et al. (2010) found that for the Local
Group a tight correlation exists between $N_{\rm S}$ and bulge
mass. Here we revisit this issue, which is necessary given that the
Local Group argument rests on only three such data points. Thus, with
this contribution we pursue to extend this correlation to galaxies
with redshifts $z<0.1$ and absolute r-band magnitudes
$M_r<-20$. Section \ref{sec:data} gives information about a pair of
catalogues used for this propose, containing respectively a list of
TDGs and morphological characterization of SDSS (Sloan Digital Sky
Survey) galaxies.  Section \ref{.corr} shows the correlation of the
number of satellites as a function of the bulge index, and we discuss
its interpretation in Section \ref{.concl}.

\section{Data}
\label{sec:data}

To test whether the above correlation between the bulge index or mass and the
number of satellite galaxies exists in other-than-Local-Group systems,
we seek to cross-correlate a catalogue of major disk galaxies with a
catalogue of TDGs. We thus use the following two catalogs:

\begin{description}
\item[Kaviraj et al. (2012, hereafter K12):] a statistical
  observational study of the TDG population in the nearby Universe
  performed by exploiting a large, homogeneous catalog of galaxy
  interactions compiled from SDSS-Data Release 6. Of all TDG-producing
  interactions 95~\% involve two spiral progenitors, while most
  remaining systems have at least one spiral progenitor. Here we
  explore how the ratio between the number of satellites and the total
  number of host or parent galaxies depends on the prominence of the
  bulge of the host galaxy. The redshifts of these TDGs (and of their
  parent galaxies) are $z\le 0.10$; their absolute magnitudes in the
  r-band, $M_r$ are between $-12$ and $-20$; their stellar masses are
  between $10^6$ and $10^{10}\,M_\odot$.  We have selected the TDGs
  with a high (rather than `unsure' or `low') confidence and flag$=1$
  (we reject those with flag$=2$, which would require further visual
  inspection of the object for it to be verified as a TDG): a total of
  508~TDGs associated with 298~different parent galaxies are obtained;
  and many of these galaxies have another galaxy associated with the
  interaction producing the given TDGs.

\item[Galaxy Zoo 2 (Willett et al. 2013):] a citizen science project
  with more than 16 million morphological classifications of 304\,122
  galaxies drawn from the SDSS-Data Release 7, with apparent r-band
  magnitude $m_r<17$, in addition to deeper images from SDSS-Stripe
  82.  Among the features described in this catalog, for spiral
  galaxies, there is a classification of the bulge intensity in
  contrast with the disc, which we associate with a `bulge index':
  0 -- no bulge; 1 -- just noticeable bulge; 2 --
  obvious bulge; 3 -- dominant bulge. Indeed, for each galaxy
  there are several votes for each classification from the different
  participants, and we use its average and root-mean-square dispersion
  defined, respectively, as
\begin{equation}
B=\frac{\sum _i\,i\times f_i}{\sum _i\,f_i},
\label{eq:B}
\end{equation}
\begin{equation}
{\rm rms}_{\rm B}=\sqrt{\frac{\sum _i\,(B-i)^2\times f_i}{\sum _i\,f_i} },
\end{equation}
where $f_i$ is the debiased \footnote{The overall effect of this bias
  is a change in observed morphology fractions as a function of
  redshift independent of any true evolution in galaxy properties;
  this is corrected for here: see Willett et al. (2013, their
  Section~3.3).} vote fraction associated to a bulge index equal to
$i$. Note that, with this definition, our bulge indices do not
  constitute ordered cardinal data, and they can be treated with the
  statistical tools of real variables.  From this catalog, we select
only the sources with redshift $0<z<0.10$, with clean flags($=1$) for
the classification as disk galaxy.  Moreover, we restrict our sample
to galaxies with $\ge 4$ votes, rms$_{\rm B}<0.5$ (i.e., we avoid
those galaxies in which there is a high dispersion of opinions about
its classification) and $-20\ge M_r>-23$ (absolute magnitudes
corrected for Galactic extinction).   This gives a total of 14\,878 galaxies.

Note that this range of absolute magnitudes is related through the
Tully-Fisher relation ($\log _{10} v_{80}=2.210-0.135(M_r+21.107)$,
where $v_{80}$ is the rotation velocity at the radius containing 80\%
of the i-band light; Pizagno et al. 2007) with the range of circular
velocities 115 km/s$\;<v_{80}<\;$295 km/s. This corresponds to
$1.2-26.3\times 10^{10}$ M$_\odot$ of baryonic mass (using the
orthogonal fit between maximum velocity and baryonic mass of
Avila-Reese et al. 2008, and assuming $v_{80}$ to be the maximum
velocity).

\end{description}

\section{Number of satellites as a function of the bulge index}
\label{.corr}

\begin{figure}
\vspace{1cm}
\centering
\includegraphics[width=8.cm]{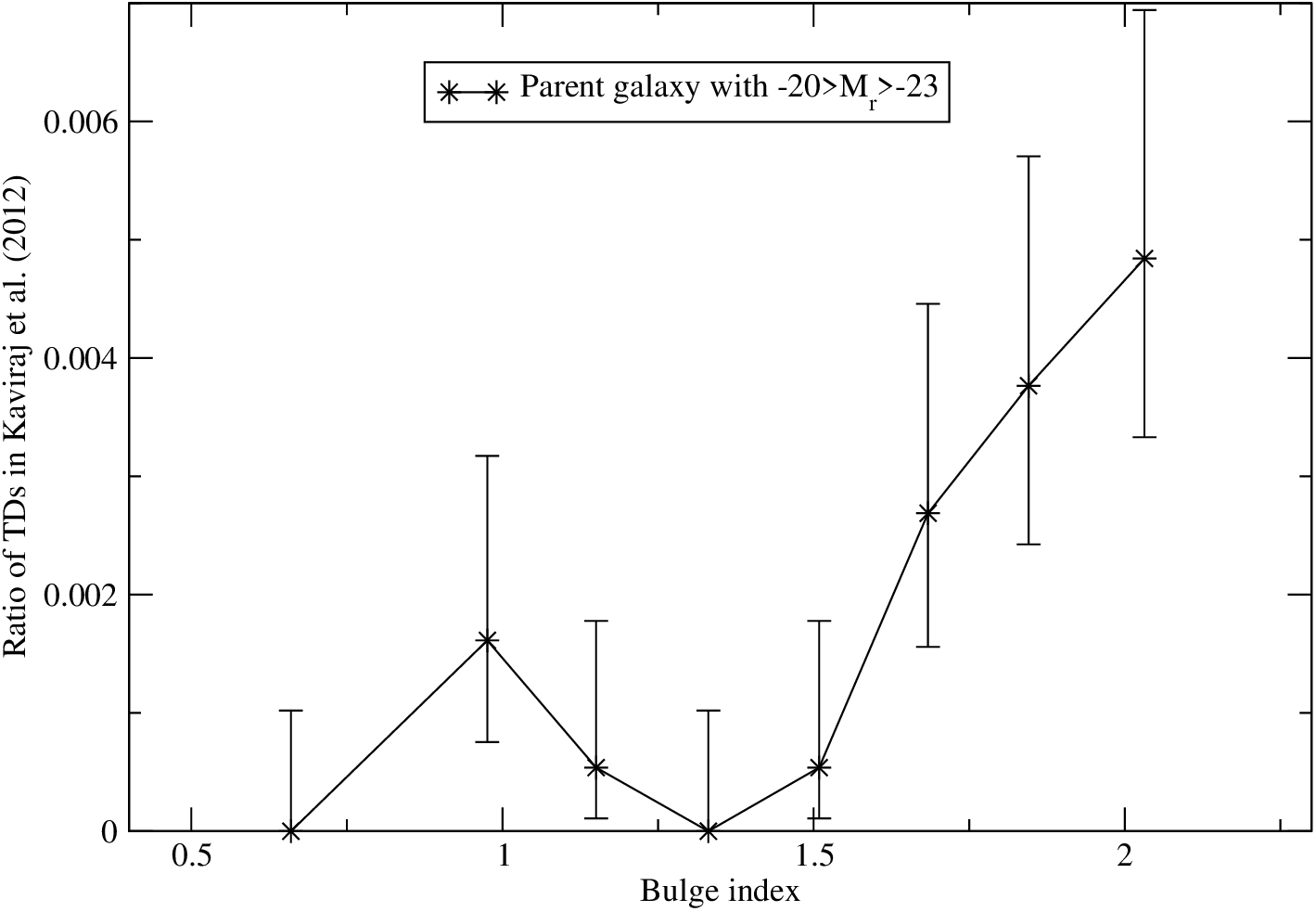}
\caption{Number of TDGs per galaxy in the selected sample in Zoo 2 as
  a function of bulge index $B$ (Eq.~\ref{eq:B}). Error bars
  stand for binomial uncertainties within 68\% C.L.}  \vspace{.2cm}
\label{Fig:ratio2}
\end{figure}

\begin{figure}
\vspace{1cm}
\centering
\includegraphics[width=8.cm]{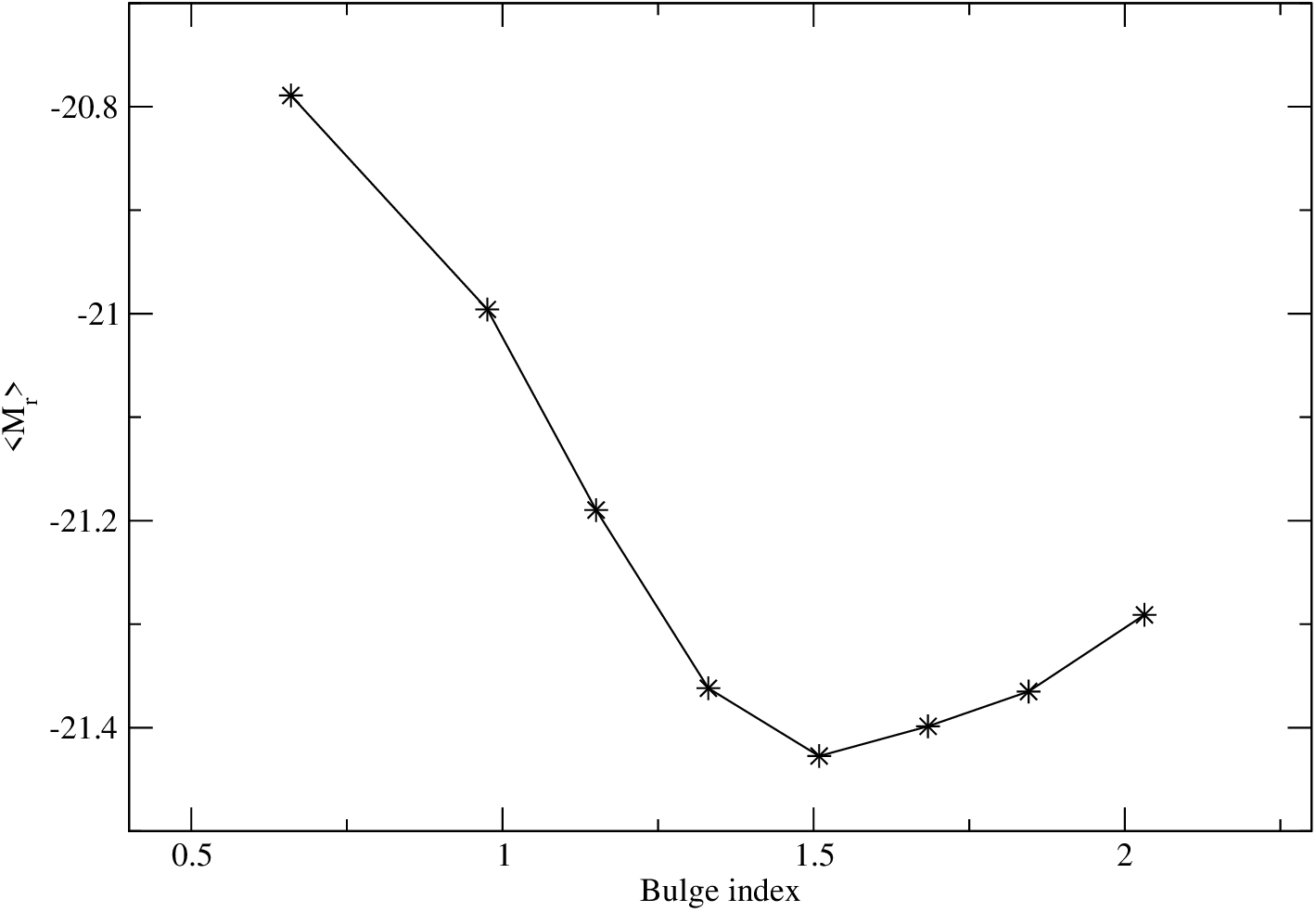}
\caption{Average absolute magnitude $M_r$ in the selected sample in
  Zoo 2 as a function of bulge index  $B$ (Eq.~\ref{eq:B}).}
\vspace{.2cm}
\label{Fig:avabsmr}
\end{figure}

\begin{figure}
\vspace{1cm}
\centering
\includegraphics[width=8.cm]{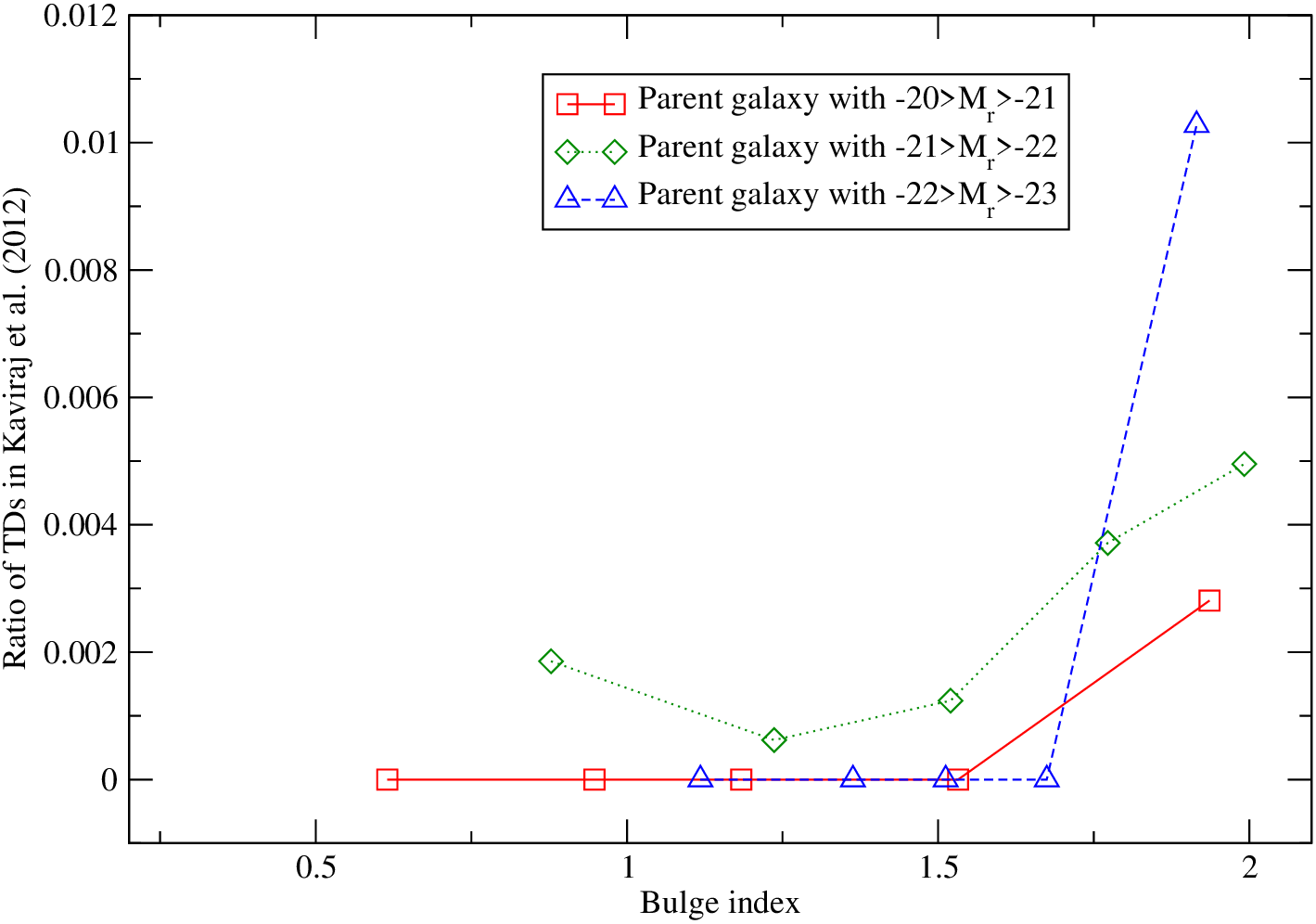}
\caption{Number of TDGs per galaxy of the selected sample in Zoo 2 as
  a function of bulge index $B$ (Eq.~\ref{eq:B})
and absolute magnitude.}
\vspace{.2cm}
\label{Fig:ratiomags}
\end{figure}

We cross-correlate the parent galaxies of TDGs with our selected
sample of Galaxy Zoo 2, allowing a maximum angular distance of~1" in
the astrometry (either with the first parent galaxy or with the
second, not with both) and a maximum difference of redshifts of
0.002. We find 26 TDGs around 16 galaxies, all of them with $z$
between 0.04 and 0.10, an average of $1.7\times 10^{-3}$ TDGs per
galaxy in the here used sample.  The distribution of the ratio between
total number of TDGs and total number of galaxies according to the
bulge index $B$ and absolute magnitudes is shown in
Figs. \ref{Fig:ratio2} and \ref{Fig:ratiomags}.  In
Fig. \ref{Fig:avabsmr} we see the variation of the average absolute
magnitude as a function of $B$.  These plots represent average ratios
for bins with the same number of galaxies.  For any absolute magnitude
range, the same trend of an increasing number of satellites with bulge
index is evident, and we note that the variation of the average
absolute magnitude is not large in our
sample. Table~\ref{Tab:TDGscorr} gives the data of the 16 galaxies
with TDGs.

\begin{table*}
\label{Tab:TDGscorr}
\caption{Galaxies from the selected sample of Galaxy Zoo 2 which are parent of some TDG.}
\begin{center}
\begin{tabular}{ccccccc}
RA, Dec. ($^\circ $, J2000)  & $z$ & $m_r$ (dered.) & $M_r$ & bulge index $B$&
rms$_{\rm B}$ & Nr. of TDGs \\ \hline
  214.9597, 44.2786 &  0.0628 &   16.15 &  -21.10 &  1.00 &  0.47 & 3 \\    
  174.7748, 10.1389 &  0.0833 &   16.04 &  -21.85 &  1.19 &  0.46 & 1 \\   
  153.6428, 26.5433 &  0.0800 &   16.52 &  -21.28 &  1.58 &  0.49 & 1 \\    
   4.6229, -0.5392 &   0.0691 &   16.25 &  -21.22 &  1.65 &  0.48 & 1 \\    
  233.7008, 58.4660 &  0.0930 &   16.52 &  -21.62 &  1.72 &  0.47 & 2 \\       
  168.4005, 28.6276 &  0.0637 &   16.32 &  -20.96 &  1.74 &  0.48 & 1 \\    
  253.0490, 21.3723 &  0.0950 &   16.57 &  -21.62 &  1.74 &  0.44 & 1 \\    
  242.8064, 52.4470 &  0.0607 &   16.63 &  -20.55 &  1.78 &  0.44 & 1 \\    
  133.9212, 57.5730 &  0.0400 &   15.47 &  -20.77 &  1.80 &  0.40 & 1 \\    
  212.2610, 3.1916  &  0.0808 &   16.46 &  -21.36 &  1.84 &  0.46 & 2 \\     
  171.1424, 30.0959 &  0.0548 &   15.85 &  -21.09 &  1.85 &  0.36 & 1 \\   
  184.4555, 35.7473 &  0.0880 &   15.37 &  -22.65 &  1.86 &  0.44 & 2 \\
  122.4296, 39.5159 &  0.0766 &   16.58 &  -21.12 &  1.97 &  0.34 & 2 \\      
  180.9550, 2.0992 &   0.0812 &   15.68 &  -22.16 &  1.99 &  0.31 & 1 \\    
  147.4184, 38.3386 &  0.0612 &   15.66 &  -21.53 &  2.15 &  0.46 & 5 \\     
  231.1698, 9.9693 &   0.0783 &   16.05 &  -21.71 &  2.35 &  0.48 & 1 \\ \hline   
\end{tabular}
\end{center}
\end{table*}

The correlation of the ratio,
\begin{equation}
r(B)\equiv \frac{{\rm Nr.\ of\ TDGs\ associated\ to\ galaxies \ with \
  bulge \ index \ B}}{{\rm Nr.\ of\ galaxies \ with \ bulge \ index \ B}}
\end{equation}
with the bulge index ($B$) is $\frac{\langle r\,B \rangle}{\langle
  r\rangle\langle B\rangle}-1=0.27\pm 0.09$. A linear fit of the type
$r=a+b\times B$ gives $a=(-2.7\pm 1.4)\times 10^{-3}$, $b=(3.2\pm
1.0)\times 10^{-3}$. Therefore, $r$ being independent of $B$ is
excluded at around 3$\sigma $. Note that this statistic is
  evaluated with the whole set of galaxies, not only within the bins
  plotted in Fig. \ref{Fig:ratio2}.

\subsection{Selection effects}

We now consider whether and how the specific
parameters used in the selection of our catalogs and their correlation
affect the results. The following issues may be considered:

\begin{itemize}
\item Dropping the constraints on the flags in the K12 catalog, we
  obtain 1644 TDGs but with lower confidence to be true TDGs, instead
  of the 508 TDGs above. Performing the analysis with these 1644 TDGs,
  83 are associated with a galaxy in the Galaxy Zoo 2 catalogue. This
  results in a more-significant correlation: $\frac{\langle r\,B
    \rangle}{\langle r\rangle\langle B\rangle}-1=0.29\pm 0.06$,
  i.e. an almost 5$\sigma $ signal. This is shown in Fig.
  \ref{Fig:ratio3}.

\begin{figure}   
\vspace{1cm}
\centering
\includegraphics[width=8.cm]{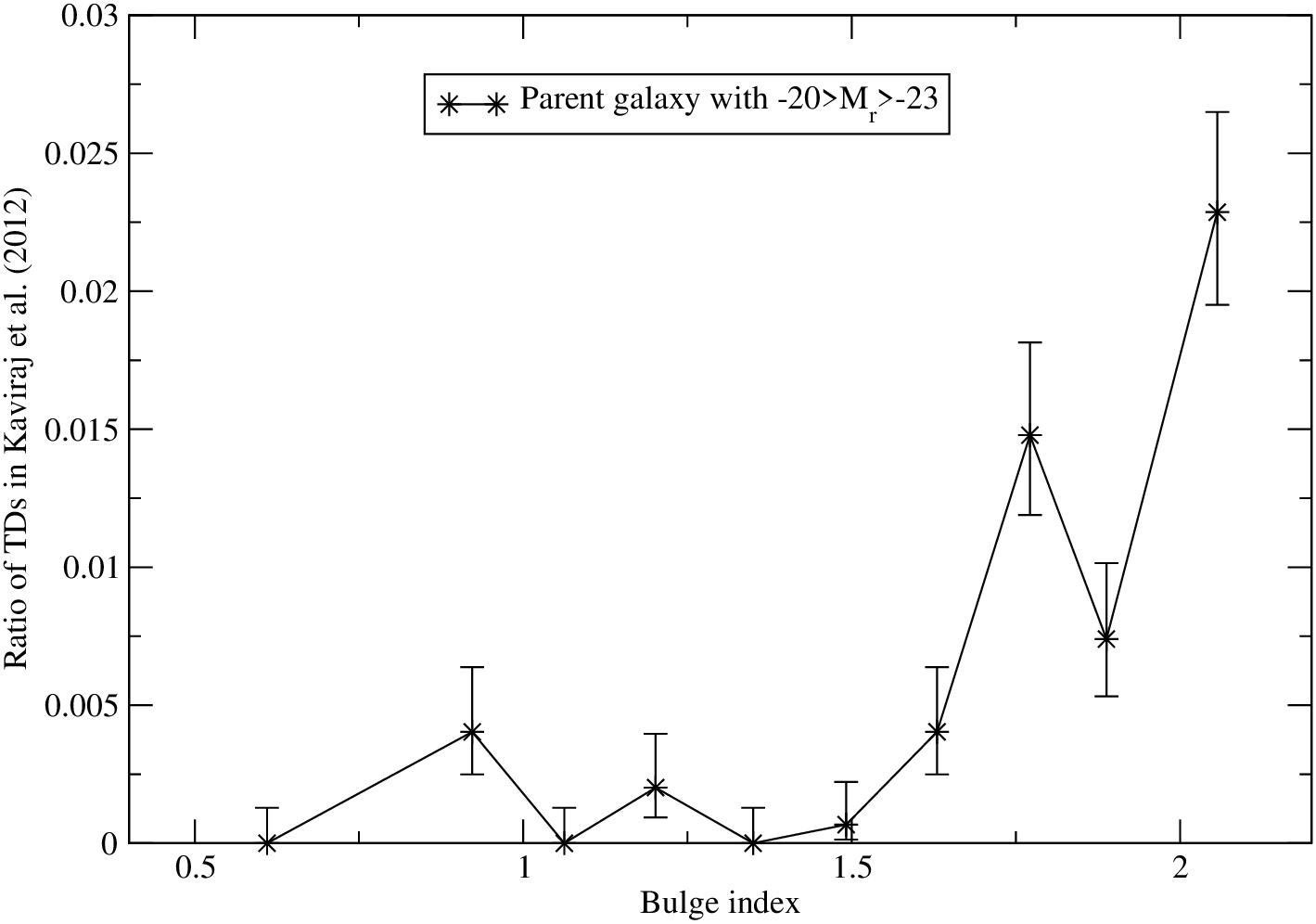}
\caption{Number of TDGs per galaxy including those with unsure or low
  confidence and flag$=2$ in the K12 catalog as a function of the
  bulge index $B$ (Eq.~\ref{eq:B}). Error bars stand for
  binomial uncertainties within 68\% C.L.}
\vspace{.2cm}
\label{Fig:ratio3}
\end{figure}

\item Dropping the constraint/flag of the clean-disc galaxy in the
  Galaxy Zoo 2 catalog, we obtain 39\,682 galaxies instead of 14\,878
  galaxies.  This analysis gives 40 TDGs associated with these
  galaxies and $\frac{\langle r\,B \rangle}{\langle r\rangle\langle
    B\rangle}-1=0.11\pm 0.07$. This is a more diluted correlation than
  previously, which is due to introducing noise in
  the host-galaxy sample  without a clear classification that they are
  disc galaxies.

\item Requiring a different amount of minimum votes, instead of four, 
for the classification of a galaxy in the Zoo~2 catalogue, we would have a value of the 
correlation $\frac{\langle r\,B \rangle}{\langle r\rangle\langle i\rangle}-1$ equal to:
$0.32\pm 0.12$ for $\ge 1$ votes (18\,972 galaxies);
$0.29\pm 0.10$ for $\ge 2$ votes (16\,903 galaxies);
$0.25\pm 0.09$ for $\ge 8$ votes (12\,519 galaxies);
$0.22\pm 0.10$ for $\ge 16$ votes (8\,912 galaxies);
$0.18\pm 0.16$ for $\ge 32$ votes (3\,688 galaxies). 
All these result in the same trend but with a slightly lower
correlation for galaxies with more votes, although being compatible
with the result of $\ge 4$ votes within the uncertainties.

\item Changing the constraint on ${\rm rms}_{\rm B}$, away from the
  above applied constraint $<0.5$, 
we find $\frac{\langle r\,B \rangle}{\langle r\rangle\langle B\rangle}-1$ equal to
$0.46\pm 0.24$ for $<0.4$ (3\,425 galaxies);
$0.15\pm 0.05$ for $<0.6$ (28\,057 galaxies);
$0.01\pm 0.03$ for $<0.8$ (44\,339 galaxies);
$-0.02\pm 0.03$ for $<1.0$ (48\,690 galaxies).
Here a clear decrease of the signal becomes evident with increasing rms.
This may be attributed to a misclassification of the galaxies with 
${\rm rms}_{\rm B}\gtrsim 0.5$; this large rms reflects indeed
that the participants of the Galaxy Zoo 2 catalogue were more in disagreement
among themselves in the classification in the included more doubtful cases.
Anyway, if we explore further the origin of the low correlation, we see that
it is mainly due to an excess of the ratio $r$ for galaxies with a
$B\lesssim 1$ (galaxies with no bulge or just a noticeable bulge). If we explore only
the region of galaxies with obvious or dominant bulge, $B>1.0$, the results
of the correlations are (note that these numbers are not directly comparable to the
previous ones because $\langle B\rangle$ is larger; pay only 
attention to the signal/noise ratio):
we get $0.22\pm 0.17$ for rms$<0.4$;
$0.21\pm 0.07$ for $<0.5$; 
$0.11\pm 0.04$ for $<0.6$;
$0.06\pm 0.03$ for $<0.8$;
$0.07\pm 0.02$ for $<1.0$.

\item The faintest galaxies in our sample are the most difficult to
  categorize in terms of them having a bulge.  Using the range of
  absolute magnitude $M_r>-20$ we get 1\,702 galaxies with correlation
  $\frac{\langle r\,B \rangle}{\langle r\rangle\langle
    B\rangle}-1=-0.12\pm 0.23$.  There is a high ratio of galaxies
  without a bulge (classified with $B<1$) but with TDGs. We attribute
  this result again to a misclassification of faint objects, all of
  them being near the limit of $m_r\approx 17$ where a visual
  inspection is more likely to fail and it is difficult to observe a
  bulge if it were present.  For the objects with $B>1$,
  $\frac{\langle r\,B \rangle}{\langle r\rangle\langle
    B\rangle}-1=0.09\pm 0.17$, which is inconclusive due to the large
  error bars given the small number statistics.  For $M_r<-23$, there
  are only 32 galaxies, and none of them have TDGs in the K12 catalog,
  so a statistical assessment of a possible correlation is not
  possible.

\item The results of the correlation do not change if we vary slightly
  the angular separation since we are using the same original source
  (SDSS) with almost the same coordinates.  The variation of the range
  of redshift to $\Delta z=$0.001, 0.003 or 0.004 instead of 0.002
  changes none of the results: the same pairs of galaxy-TDG are
  detected.

\end{itemize}

Summing up, our result is robust against the change of parameters used
to estimate the relationship between the ratio of TDGs and the bulge
index, giving a correlation with a significance of up to 5$\sigma $ in
the best of the cases. Five sigma is a very improbable configuration
(probability lower than one in a million) so we do not think this is
by chance due to a posteriori statistics.  We do not
    have a continuous variable (the number of TDGs) from which we
  have taken the most convenient value in order to
    improve the signal/noise to our advantage, but
  instead we have a discrete variable (the flag) by
  which either we take the whole sample or we exclude those TDGs with
  flag=2; there are only these two options. And if, just including
  flag=2 TDGs, we get a significance several orders of magnitude
  higher, this cannot be due to fine tuning, but is due to the
  detection of a real signal which is amplified by the
  increase of the number of TDGs. In any case, the result with the
most conservative sample of secure TDGs gives a correlation between
the number of TDGs and the bulge index which has a significance of
$3\,\sigma$. The  above discussed departures from
this significance  when the criteria are varied are
due to a high contamination of galaxies with possible wrong
classifications.

\section{Discussion and conclusions}
\label{.concl}

As discussed in the introduction, two competing cosmological
frameworks exist for the emergence of galaxies, the dark-matter based
standard models and the generalized-gravity models without cold or
warm dark matter. They differ by dynamical friction on the expansive
and massive dark matter halos not acting in the generalized-gravity
models, such that the formation and growth of galaxies differs
significantly. In the dark-matter-based models, each major galaxy has
many dark-matter-dominated satellite galaxies which have independent
infall histories and are captured around the host galaxy through
dynamical friction. In the generalized-gravity models companion
galaxies are either on hyperbolic fly-bys or are TDGs which form,
together with populations of star clusters, phase-space correlated
populations around the major hosts. The observational evidence
strongly favors the latter models, given the small number of satellite
galaxies observed around major galaxies and given their ubiquitous
phase-space correlations. Another implication of the
generalized-gravity models is that galaxies rarely merge but they
nevertheless interact. Interactions form bulges and TDGs. A strong
correlation between the bulge mass and the number of faint satellite
galaxies has been noted to exist for Local Group galaxies. Here we
revisit this issue by considering a large ensemble of host galaxies
which we cross-correlate with catalogues of TDGs. The results suggest
a strong correlation exists between the bulge index and the number of
TDG companions. This supports the result obtained for the Local Group
and therefore also the generalized-gravity models.

Can the SMoC explain in some way this correlation of the number of
satellites with the bulge index?  In principle, we see only one
possibility: that the bulge/disk ratio is correlated with the halo
mass, i.e. that the formation of a classical bulge or pseudobulge/bar
depends on the halo mass.  The bulge may be a classical bulge,
i.e. with a stellar orbit distribution typical of elliptical galaxies
and formed at the early stages of the host galaxy's life, or it may be
a pseudo-bulge, or a long bar, or a misaligned triaxial bulge and a
long bar may both be present in spiral galaxies (Comp\`ere et
al. 2014). In any of the cases, the situation is similar.

Later galaxy types (higher Hubble stage) have an observed smaller
bulge index (Simien \& de Vaucouleurs 1986) and have lower
luminosities on average (Graham \& Worley 2008). Therefore, one may
suspect that the average mass of the host-galaxies in our sample
increases with the bulge index. But there is an anticorrelation of the
fraction of satellites with the luminosity or with the stellar mass
(Velander et al. 2014), so a higher luminosity or stellar mass of
galaxies with prominent bulges would give fewer satellites. This is
the opposite of what we see. Nonetheless, the variation of the average
absolute magnitude in our sample is not large (see
Fig. \ref{Fig:avabsmr}) and, even if we subdivide the sample into
smaller ranges of absolute magnitude, the trend of more satellites
with higher bulge index is retained (see Fig.
\ref{Fig:ratiomags}). Moreover, the ratio of the dark-to-luminous mass
in galaxies is either higher with higher Hubble stage (Tinsley 1981)
or it is uniform (Jablonka \& Arimoto 1992).  But, the Snyder et
al. (2015) results obtained using the ``Illustris'' simulations to
represent the SMoC, point out that bulge-dominated galaxies should
have a higher ratio of dark-to-stellar mass, so one may thus note that
stellar mass and luminosity are not proportional.  Therefore, assuming
an approximate constant luminosity, the total dynamical mass is either
lower or roughly the same with larger bulge index. This is the
opposite of what would be needed to explain the correlation discovered
here within the dark matter scenario.  With respect to the existence
of long bars, data on rotation curves do not show a clear trend
towards higher halo masses for barred galaxies (L\'opez-Corredoira
2007). Hence, given the above we do not see how the SMoC can explain
our results.  Further research is needed to explore the possibility
that the bulge index in the SMoC might correlate with other quantities
which also influence the number of TDGs (e.g., measures of the
assembly history of the halo).  What is evident from our analysis is
that the available data favor the Milgromian model, which gives a
simple and direct explanation of the data, over the SMoC within which
one needs to explore complex ad hoc modifications of the application
of the theory in order to explain something which was not predicted a
priori.

To put this finding on a more secure footing a dedicated observational
survey is required as follows (Kroupa 2015): A catalogue of disk
galaxies with similar circular velocities but different bulge masses
are needed. Each of these disk galaxies needs to be surveyed over
regions with radii of 150 to 250~kpc to seek faint dSph-like satellite
galaxies. Such a survey will allow a quantification of a possible
bulge-mass vs number of satellite galaxy correlation to seek
confirmation or rejection of the correlation found in the Local Group
and with this study.  The implications of the existence or absence of
such a correlation would pose important empirical constraints on
fundamental physics because it is directly related to the question of
whether dark matter particles exist, or whether the standard model of
particle physics remains the best current description of all existing
particles.

\acknowledgments We thank Sugata Kaviraj for kindly providing the
catalog of tidal dwarfs published in Kaviraj et al. (2012)
  and Marcel Pawlowski for useful comments. An anonymous
  referee is acknowledged for requesting a more detailed discussion on
  satellite planes.  MLC was supported by the grant AYA2012-33211 of
the Spanish Ministry of Economy and Competitiveness (MINECO).  Funding
for the SDSS-II has been provided by the Alfred P. Sloan Foundation,
the Participating Institutions, the National Science Foundation, the
U.S. Department of Energy, the National Aeronautics and Space
Administration, the Japanese Monbukagakusho, the Max Planck Society,
and the Higher Education Funding Council for England. The SDSS Web
Site is http://www.sdss.org/.  The SDSS is managed by the
Astrophysical Research Consortium for the Participating
Institutions. The Participating Institutions are the American Museum
of Natural History, Astrophysical Institute Potsdam, University of
Basel, University of Cambridge, Case Western Reserve University,
University of Chicago, Drexel University, Fermilab, the Institute for
Advanced Study, the Japan Participation Group, Johns Hopkins
University, the Joint Institute for Nuclear Astrophysics, the Kavli
Institute for Particle Astrophysics and Cosmology, the Korean
Scientist Group, the Chinese Academy of Sciences (LAMOST), Los Alamos
National Laboratory, the Max-Planck-Institute for Astronomy (MPIA),
the Max-Planck-Institute for Astrophysics (MPA), New Mexico State
University, Ohio State University, University of Pittsburgh,
University of Portsmouth, Princeton University, the United States
Naval Observatory, and the University of Washington.

\end{document}